\documentclass[a4paper]{JHEP3}
\usepackage{epsfig}
\title{Locality and Statistical Error Reduction on Correlation Functions}
\author{Harvey B. Meyer\\Theoretical Physics, University of Oxford, 
1 Keble Road,\\ Oxford, OX1 3NP, United Kingdom\\Email: \email{meyer@thphys.ox.ac.uk}}
\abstract{We propose a multilevel Monte-Carlo scheme, applicable to local actions, which is expected to reduce statistical errors on correlation functions.
We give general arguments to show how the efficiency and parameters of the
 algorithm are determined by the low-energy spectrum. As an application,
we  measure the euclidean-time correlation of pairs of Wilson loops 
in $SU(3)$ pure gauge theory with constant relative errors. 
In this case the ratio of the new method's 
efficiency to the standard one increases as $e^{m_0t/2}$, 
where $m_0$ is the mass gap and $t$ the time separation.}
\preprint{hep-lat/0209145}

\newcommand{\be}{\begin{equation}}
\newcommand{\ee}{\end{equation}}
\newcommand{\ba}{\begin{eqnarray}}
\newcommand{\ea}{\end{eqnarray}}
\newcommand{\bi}{\begin{itemize}}
\newcommand{\ei}{\end{itemize}}
\newcommand{\tr}{{\rm Tr\,}}
\newcommand{\re}{\mathop{\rm Re}}

\newcommand{\<}{\langle} 
\renewcommand{\>}{\rangle}

\newcommand{\la}[1]{\label{#1}}
\begin{document}
%Oxford-OUTP-02-??\\
%\setcounter{footnote}{0}
%%%%%%%%%%%%%%%%%%%%%%%%%%%%%%%%%%%%%%%%%%%%%%%%%%%%%%%%%%%%%%%%%%%%%%%
\section{Introduction}
In equilibrium statistical mechanics and  quantum field theory, 
the physical information on the theory is encoded in $n$-point functions.
The short-range nature of interactions in the former and the causality 
requirement in the latter case lead to the property of locality of the 
Hamiltonian (resp.  action). While certain lower-dimensional
systems have been solved analytically, Monte-Carlo simulations have become
an invaluable tool in the study of interacting theories such as non-abelian
gauge theories. In particular, the properties of the spectrum are extracted
from numerically calculated 2-point functions in the euclidean formulation:
\[ C(t)\equiv\langle{\cal O}^\dagger(t) {\cal O}(t=0)\rangle =\sum_n \left|\langle 0|
 {\cal O}|n\rangle\right|^2e^{-E_nt}\]
This formula follows directly from the insertion of the complete set of 
energy eigenstates $H|n\rangle=E_n|n\rangle$. 
In theories with a mass gap, the exponential 
decay of each term singles out the lightest state compatible
 with the symmetry of the operator, thus enabling us to extract the
 low-lying spectrum of the theory. However, it is precisely this decay that 
makes the 2-point function numerically difficult to compute at large $t$.
Indeed, standard algorithms keep the absolute error roughly constant, 
so that the error on the local effective mass 
\[am_{\mathrm{eff}}(t+\frac{a}{2})=\log{\frac{C(t)}{C(t+a)}}\]
increases exponentially with the time separation. For that reason, 
it would be highly desirable to have a more efficient
method to compute correlations functions at large time separation $t$. The
task amounts to reduce uncorrelated fluctuations between the two time slices
separated by euclidean time $t$.

In the context of non-abelian gauge theories, 
two types of techniques have proved useful: the first aims essentially at 
increasing the ``signal''. The widely used ``fuzzing'' algorithms, such as
smearing \cite{smear} and blocking \cite{block}, increase the overlap
 onto the lightest states. As the lattice spacing $a$ is made much smaller
 than the correlation length $\xi$~(\footnote{e.g. defined by
 $1/\sqrt{\sigma}$, where $\sigma$ is the fundamental string tension}), 
the number
of smearing/blocking steps as well as their weights have to be increased in
order to maintain the quality of the overlap.

The other type of technique aims at reducing the ``noise'', that is, the 
variance of the quantity being evaluated. For instance, when computing a 
Polyakov loop with the multihit technique~\cite{multihit}, 
a link is replaced by its thermal
average under fixed neighbouring links. Last year, 
L\"uscher and Weisz~\cite{luwei1} demonstrated how to make even better 
use of the locality of the theory to exponentially improve the efficiency of
 the algorithm that computes Polyakov loop correlators. 
In that method, pairs of time-like links
are averaged over in time slices of increasing width
 with fixed boundary conditions 
before they are multiplied together to form the loops.
In this paper, we present a ``noise reduction'' method that exploits the 
locality property in a similar way. It has the advantage of being compatible 
with the popular link-smearing and -blocking: one can cumulate the advantages
of both types of techniques. Indeed, while the fuzzing algorithms also
help reduce short wavelength fluctuations inside a time-slice, till now
no attempt has been made to tame the noise induced by fluctuations 
appearing in neighbouring time-slices. 
Our method aims at averaging out the latter by performing 
additional sweeps with fixed time-slices as boundary conditions; 
as the continuum is approached, the volume
over which the average is performed ought to be kept fixed in physical units 
to maintain the efficiency of the algorithm.
Finally, we note that the idea is very general and is expected
 to be applicable in other types of theories.

\paragraph{Outline.-} We first present the idea in its full generality, 
without reference to the specific form of the action or to the quantity
being computed. We next formulate a multilevel scheme for the case of a 
2-pt function and point out how the efficiency and parameters of the 
algorithm are determined by the low-lying spectrum of the theory.
We finally apply this algorithm to the case of (3+1) $SU(3)$
gauge theory.

%%%%%%%%%%%%%%%%%%%%%%%%%%%%%%%%%%%
\section{Locality, hierarchical formula and  multilevel algorithm}
%%%%%%%%%%%%%%%%%%%%%%%%%%%%%%%%%%%
\subsection{Locality}
%%%%%%%%%%%%%%%%%%
The locality property of most studied actions allows us to  derive
 an interesting way of computing correlation functions. 
First we give a general, 'topological' definition of locality in continuum
field theories. For simplicity we shall use a symbolic notation; 
 if ${\cal C}$ denotes a configuration, let 
${\cal X}$, ${\cal Y}$ and ${\cal A}$ be 
mutually disjoint subsets of ${\cal C}$. If $\Omega_X$,  $\Omega_Y$ and 
$\Omega_A$ are their (disjoint) supports on the (continuous) space-time
 manifold $\cal M$, suppose furthermore that any continuous path 
$\gamma: I \rightarrow {\cal M}$ joining  $\Omega_X$ and $\Omega_Y$
(i.e. $\gamma(I)\cap \Omega_X\neq \emptyset$ and  
$\gamma(I)\cap \Omega_Y\neq \emptyset$)
passes through $\Omega_A$ (i.e. $\gamma(I) \cap \Omega_A \neq \emptyset$).
 See fig. (\ref{topol}).
%%%%%%%%%%%%%%%%%%%%%%%%%%%%%%%%%%%%%%%%%%%%%%%%%%%%%%%%%%%%%%%%%%%%%%%%%%%%%
\FIGURE[htb]{

\centerline{\begin{minipage}[c]{8cm}
    \psfig{file=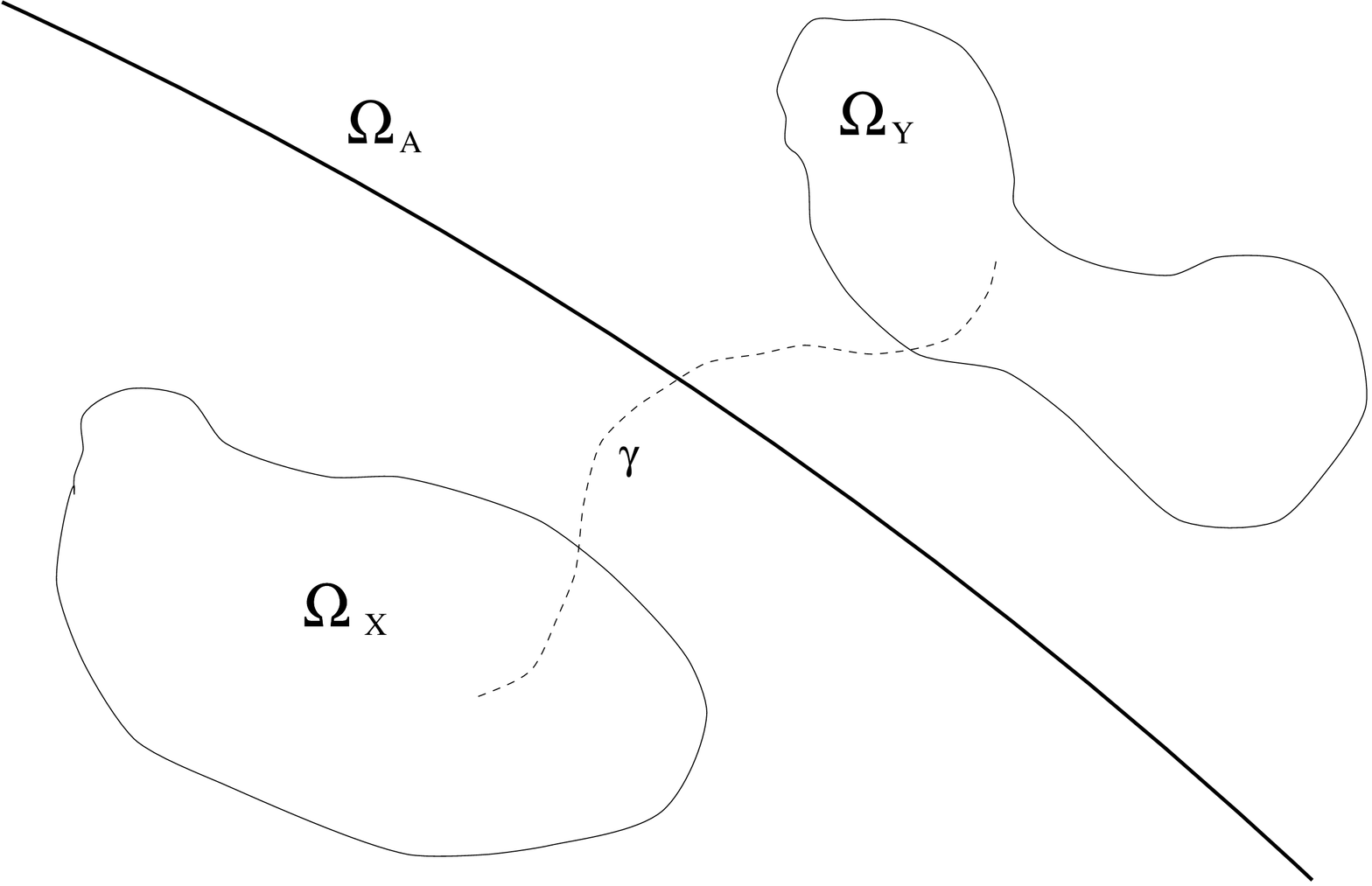,angle=270,width=6cm}
    \end{minipage}}

\vspace*{0.5cm}

\caption[a]{The sets $\Omega_X$,  $\Omega_Y$ and  $\Omega_A$ on the 
space-time manifold.}

\la{topol}
}
%%%%%%%%%%%%%%%%%%%%%%%%%%%%%%%%%%%%%%%%%%%%%%%%%%%%%%%%%%%%%%%%%%%%%%%%%%%%%

  The theory with probability distribution 
$p({\cal C})$ is \emph{local} if there exists functionals $p_{\cal A}$ and
$\tilde p_{\cal A}$ such that, for any  setup with this topology, 
\be p({\cal X}, {\cal Y})= \sum_{\cal A} p({\cal A})~p_{\cal A}({\cal X})~
 \tilde p_{\cal A}({\cal Y})\la{ppp}
\ee
That is, ``${\cal X}$ and ${\cal Y}$ influence each other
only through ${\cal A}$''.
This condition is obviously satisfied by continuum euclidean 
field theories whose 
Lagrangian density contains a finite number of derivatives. With a suitable 
notion of ``continuity'' of the path $\gamma$, one can extend this definition
to lattice actions. For instance, the Wilson action 
%(\cite{wils})
 is also local in this sense,
 but note that  $\Omega_{\cal X}$ and $\Omega_{\cal Y}$ 
must be  separated by more than one lattice spacing in order to realise
 the setup in the first place.
%%%%%%%%%%%%%%
\subsection{Hierarchical formula}
%%%%%%%%%%%%%%
As a consequence, for two operators ${\cal O}_x$ and ${\cal O}_y$,
functionals of ${\cal X}$ and ${\cal Y}$ respectively, we have
\be \< {\cal O}_x {\cal O}_y \> 
\equiv \sum_{{\cal C}} {\cal O}_x({\cal C}) {\cal O}_y({\cal C}) p({\cal C})
=\sum_{\cal A} p({\cal A})~ \< {\cal O}_x \>_A ~ \< {\cal O}_y \>_A
\la{form}
\ee
where
\ba
\< {\cal O}_{x} \>_A &=& \sum_{\cal X} p_A({\cal X}) 
~{\cal O}_x({\cal X})\nonumber\\
\< {\cal O}_{y} \>_A &=& \sum_{\cal Y} \tilde p_A({\cal Y})
~ {\cal O}_y({\cal Y}) \label{subav}
\ea
are the average values of the operators at a fixed value of $\cal A$. Thus
the averaging process factorises into an average \emph{at} fixed ``boundary 
conditions'' and an average \emph{over} these boundary conditions. There
are several ways in which this factorisation can be  iterated:
first, if the operator ${\cal O}_x\equiv{\cal O}_{x_1} ~{\cal O}_{x_2}$ 
itself factorises, the decomposition can be carried out also at this level,
where $p_A$ now plays the role of $p$. This means that the decomposition
(\ref{form}) allows us to treat the general $n$-point functions in the
 same way as the $n=2$ case that we shall investigate in more detail:
each factor can be averaged over separately.

There is another way the decomposition can be iterated: we can in turn write
  $\<{\cal O}_x\>_A$ and $\<{\cal O}_y\>_A$ as factorised averages over
yet smaller subspaces, thus obtaining a nested expression for the 
correlation function. A three-level version of (\ref{form}) would be
\ba
\< {\cal O}_x {\cal O}_y \>& =&\sum_{\cal A} p({\cal A})~\times~
\sum_{{\cal A}_1} p_A({\cal A}_1)~\sum_{{\cal A}_2} p_{A_1}({\cal A}_2)
\<{\cal O}_x\>_{{\cal A}_2}~ \times \nonumber\\
&&\sum_{\tilde{\cal A}_1} \tilde p_A(\tilde{\cal A}_1)~
\sum_{\tilde{\cal A}_2} \tilde p_{\tilde A_1}(\tilde{\cal A}_2)
\<{\cal O}_y\>_{\tilde{\cal A}_2}\la{hifo}
\ea
This type of formula is the basis of our multilevel algorithm for the 2-point 
function. 

%%%%%%%%%%%%%%%%%%%%%%%%%%%
\subsection{Multilevel algorithm for the 2-point function}
%%%%%%%%%%%%%%%%%%%%%%%%%%%
The hierarchical formula (\ref{hifo}) is completely analogous
to the expression derived in \cite{luwei1} in the case of the Polyakov loop,
where it was also proved that it can be realised in a Monte-Carlo simulation
by generating configurations in the usual way, then keeping the subset
$\cal A$ fixed and  updating the regions $\cal X$ and $\cal Y$.
Suppose we do $N$ updates of the boundary
and $n$ measurements of the operators for each of these updates. We are thus
 performing $N\cdot n$ updates\footnote{An update in general consists of
 several sweeps.}. But because the two sums in (\ref{form})
are factorised, this in effect achieves $N n^2$ measurements.
 As long as 
\begin{itemize}
\item the latter are independent;
\item  that the
fluctuations on the boundary $\Omega_{\cal A}$ have a much smaller influence
than those occurring in $\Omega_{\cal X}$ and $\Omega_{\cal Y}$;
\item that no phase transition occurs \cite{luwei1}
due to the small volume and the boundary conditions;
\end{itemize}
error bars reduce with the computer time $\tau$ like $1/\tau$ rather than
 $1/\sqrt{\tau}~$: to divide the error by two,  we  double $n$. 
The fluctuations
of the boundary are only reduced in the usual $1/\sqrt{\tau}$ regime.
Thus, for a fixed overall computer time, 
one should tune parameters of the multilevel algorithm
 so that the fluctuations in $\cal X$ and $\cal Y$ 
are reduced down to the level of those coming from $\cal A$. 

The one-level setup
we shall use in practise is illustrated in fig. (\ref{timesl}):
$\Omega_{\cal X}$, $\Omega_{\cal Y}$ and $\Omega_{\cal A}$ are time-slices.
%%%%%%%%%%%%%%%%%%%%%%%%%%%%%%%%%%%%%%%%%%%%%%%%%%%%%%%%%%%%%%%%%%%%%%%%%%%%%
\FIGURE[tb]{

\centerline{\begin{minipage}[c]{8cm}
    \psfig{file=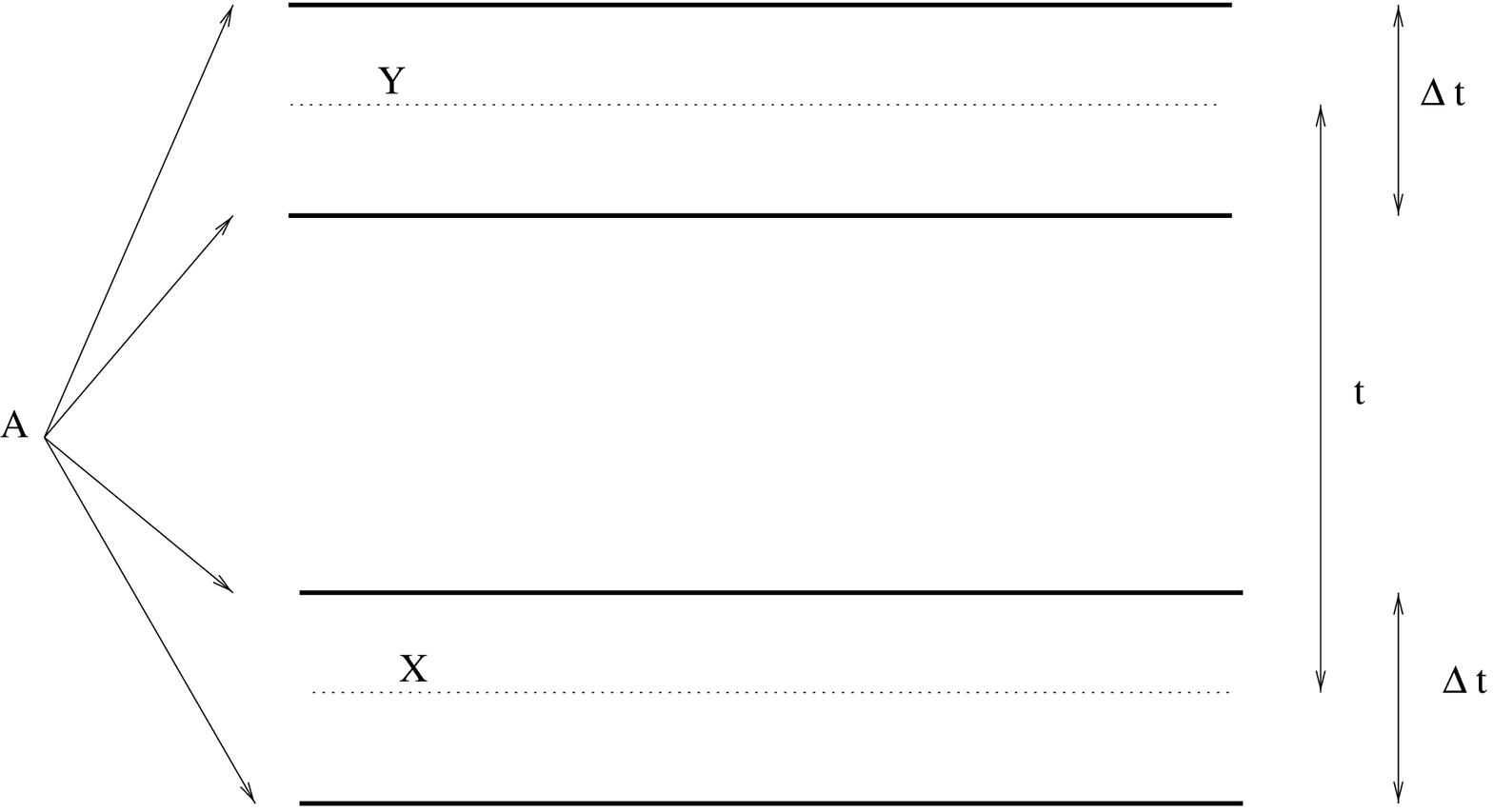,angle=270,width=6cm}
    \end{minipage}}

\vspace*{0.5cm}

\caption[a]{Our choice of $\Omega_{\cal X}$, $\Omega_{\cal Y}$ and $\Omega_{\cal A}$ to implement the hierarchical formula.}

\la{timesl}
}
%%%%%%%%%%%%%%%%%%%%%%%%%%%%%%%%%%%%%%%%%%%%%%%%%%%%%%%%%%%%%%%%%%%%%%%%%%%%%
An indication of how many 
``submeasurements'' should be chosen at each level
is given by the following consideration:
if we measure an operator located in the middle of a time-block 
 (of width $\Delta t$) bounded by $\cal A$, then
any fluctuation occurring on $\cal A$ can be decomposed on the basis which
diagonalises the Hamiltonian of the theory. If  the lowest-lying state
 compatible with the symmetry of the operator being measured has  mass $m_0$
and $\Delta t > 1/m_0$, that state will act as the main 'carrier' of
 the fluctuation, so that it will induce a fluctuation of relative 
magnitude $e^{-m_0\Delta t/2}$ on the time-slice where the operator is 
measured (see fig. 1); indeed the propagation of fluctuations is damped
 exponentially in a system that develops a mass gap.
Thus it is worth performing roughly $n\sim e^{m_0\Delta t}$ 
measurements\footnote{These may be  nested, in which case it is the total
number of measurements under fixed boundary $\cal A$ that is meant here.}
  under fixed boundary $\cal A$ in order to reduce the 
fluctuations coming from  $\cal X$ down to the level of those of $\cal A$.
In fact, this estimate is an upper bound, because the vacuum state
under the fixed boundary conditions could lie at a higher energy level
than the full-lattice vacuum.
Finally, if zero modes are present, we expect a power law 
$n\propto (\Delta t)^\eta$.

At any rate, we need to optimise
the parameters of the multilevel Monte-Carlo algorithm ``experimentally'', 
but we shall see in a practical example 
that this order of magnitude estimate is in qualitative
agreement with the outcome of the optimisation procedure.

This simple argument also shows that the  purpose of the 
multilevel scheme is to reduce fluctuations occurring
at all separations (from the time-slice where the operator is measured) 
ranging from 0 to $\Delta t/2$ with an appropriate number of 
updates, in order to reduce their influence down to the level of the out-most
boundary, which is then averaged over in the standard way. 

\subsection{Error reduction\label{errestim}}
Suppose the theory has a mass gap and that for a given $t$, 
the correlation function $C(t)\sim e^{-mt}$ 
is determined with equal amounts of computer time with the standard algorithm 
and the multilevel one. 
If we now want to compute $C(2t)$ with the same \emph{relative} precision
with the former, we need to increase the number of measurements by a factor
$e^{2mt}$. With the multilevel algorithm however, we would e.g. introduce an
extra level of nesting, so as to multiply the number of submeasurements 
by $e^{mt}$, as explained in the preceding section. Thus in this situation,
the gain in computer time of this method is a factor $e^{mt}$; 
turned the other way, it achieves
an error reduction $\propto e^{-mt/2}$ compared to the standard algorithm
\footnote{Note however that the computer effort is still increasing 
exponentially with $t$.}. Since the quantity determining the error reduction
is the product $mt$, the error reduction is in first approximation 
independent of $\beta$, as long as we measure the correlation function
at fixed $t$ in physical units. It would be inefficient to perform
sweeps over time-blocks that are much thinner than the physical length scale: 
none of the three conditions highlighted in the preceding section is likely
to be satisfied.

If we compare this to the error reduction analysis in \cite{luwei1},
we note a  ``conservation of difficulty'' law: in Wilson loop calculations, 
the exponent in the error reduction factor is the area of the loop, while
here it is the time separation of the two operators. 

%%%%%%%%%%%%%%%%%%%%%%%%%
\section{The algorithm in practise}
%%%%%%%%%%%%%%%%%%%%%%%%%
%%%%%%%%%%%%%%%
\subsection{Preliminaries}
%%%%%%%%%%%%%%%%%
We consider pure $SU(3)$ lattice gauge theory in
 $(3+1)$ dimensions and we use the standard Wilson action
\begin{equation}
S[U]=\beta \sum_{x,\mu<\nu} (1-\frac{1}{3}\re\tr{P_{\mu\nu}}) \la{wils}
\end{equation}
where $U$ are the link variables, $P_{\mu\nu}$ is the plaquette and
 $\beta=\frac{6}{g_0^2}$.

We define  zero-momentum operators as a sum over a time-slice 
of Wilson loop traces:
\be
{\cal O}(t)= \sum_{\vec{x}} \tr{W(\vec x, t)}
\ee
We are interested in measuring correlation functions of these operators:
\be
\< {\cal O}^\dagger(0)~{\cal O}(t) \> \equiv \frac{1}{{\cal Z}} \int D[U]
~{\cal O}^\dagger(0){\cal O}(t)~ e^{-S[U]}, \quad D[U]=\prod_{x,\mu}dU(x,\mu)
\la{cf}
\ee
The main physical information we extract from this numerical measurement
 is the mass of the lightest
state compatible with the symmetry of ${\cal O}$, by fitting the exponential
decay of the correlation function. Numerically the path integral is evaluated
by  updating the lattice configuration according to the Boltzmann weight
 and evaluating the operators at each of these updates.
Here we calculate the correlation function of two operators carrying the 
continuum quantum numbers $0^{++}$ and $2^{++}$, 
constructed with $4\times 2$ rectangular Wilson loops. 
The simulation is done at $\beta=5.70$ on an $8^4$ lattice. 
For the update we use a mixture of heat-bath~\cite{kenpen}  
and over-relaxation~\cite{adler} in the ratio 1:4, 
both applied to three SU(2) subgroups~\cite{hb} .
Finally, as was noted in~\cite{luwei1}, the most elegant way to implement a 
multilevel Monte-Carlo program is to use a recursive function.
%($8a\simeq 1.36$fm)

%%%%%%%%%%%%%%%%%%%%%%%%%%%%%%%%%%%
\subsection{Averages under fixed boundary conditions}
%%%%%%%%%%%%%%%%%%%%%%%%%%%%%%%%%%%
%%%%%%%%%%%%%%%%%%%%%%%%%%%%%%%%%%%%%%%%%%%%%%%%%%%%%%%%%%%%%%%%%%%%%%%%%%%%%
\FIGURE[htb]{

\centerline{\begin{minipage}[c]{10cm}
    \psfig{file=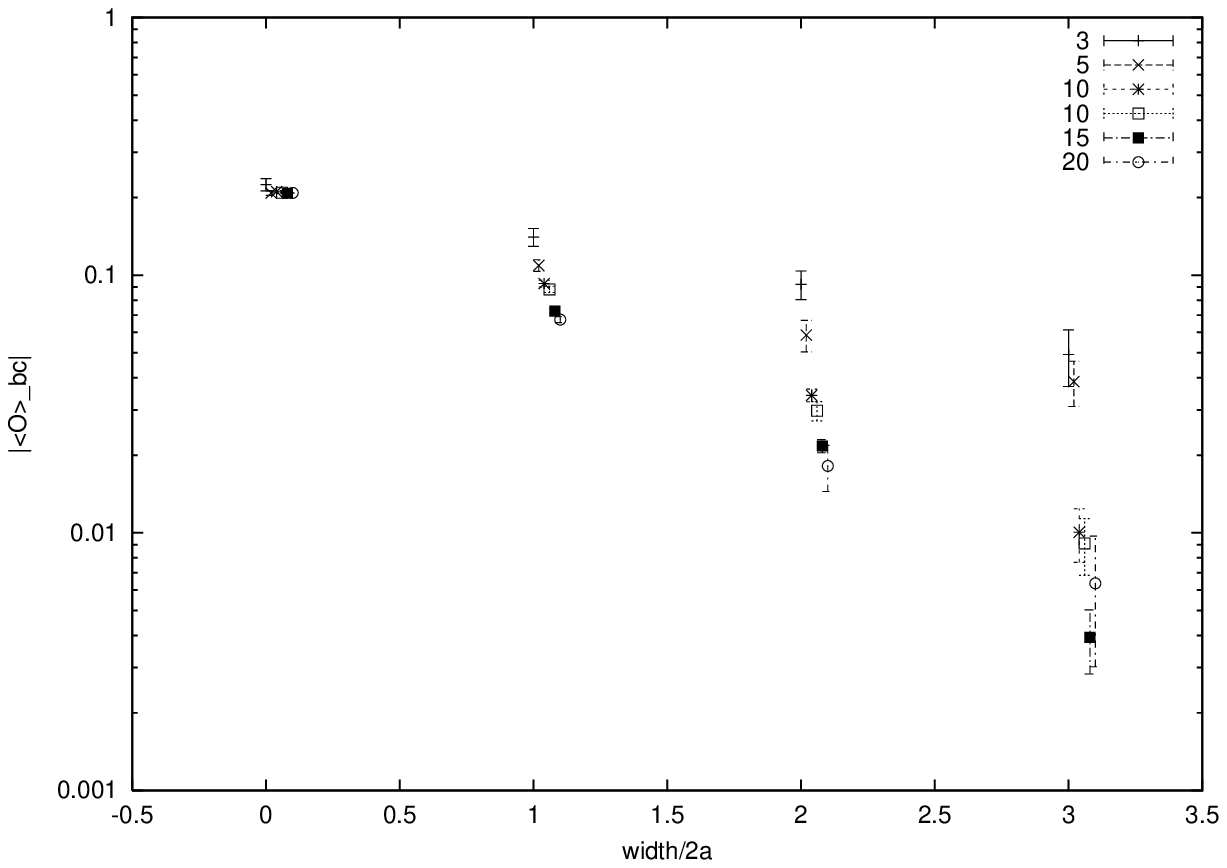,width=10cm}
    \end{minipage}}
\vspace{0.5cm}
\centerline{\begin{minipage}[c]{10cm}
    \psfig{file=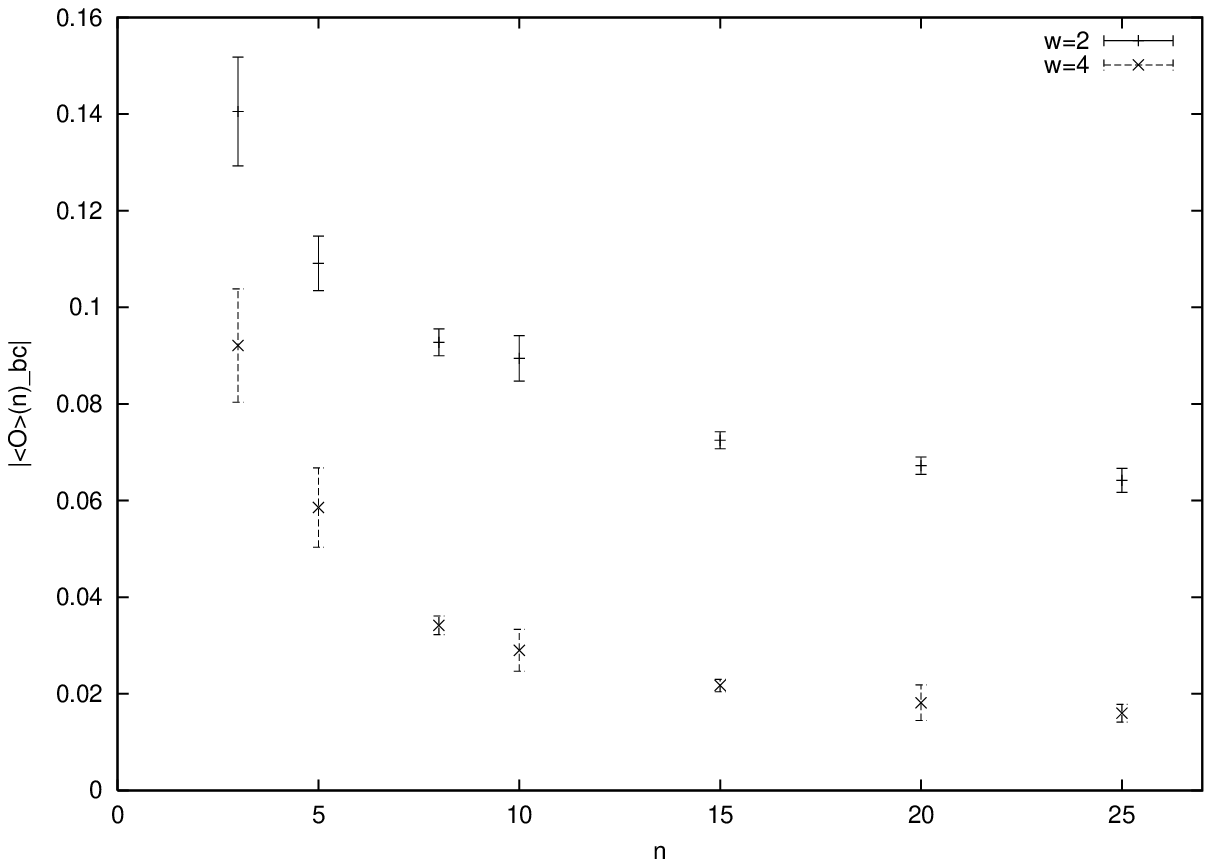,width=10cm}
    \end{minipage}}

\caption[a]{``Error'' on the average value of the $2^{++}$ operator ($4\times2$ rectangles), as a 
function of the width of the time-block (top) and the number of measurements
(bottom).}

\la{stud}
}
%%%%%%%%%%%%%%%%%%%%%%%%%%%%%%%%%%%%%%%%%%%%%%%%%%%%%%%%%%%%%%%%%%%%%%%%%%%%%
To illustrate our arguments on the way an error reduction is achieved,
we  measure the change of the average value of an  
operator carrying quantum numbers $2^{++}$ --
whose expectation value in the full lattice is known to be exactly zero --
 due to fluctuations on the boundary and the finite number of measurements
$n$ performed:
\[G_n(\Delta t)= \sum_{\cal A} p({\cal A}) ~|\<{\cal O}^{2^{++}}\>^{(n)}_{\cal A}|\]
 Fig. (\ref{stud}, top) shows the dependence on
 the width $\Delta t$ of $G$, for
several fixed values of $n$. The data was obtained by nesting averages on 
increasing widths of the time-slices. 
Here, to reduce the number 
of parameters, we set $n$ to be the same for all time-slices. 
Clearly, the data confirms that 
$G(\Delta t)$ vanishes exponentially, as anticipated (this
\emph{a posteriori} justifies the choice of keeping $n$ constant).
The exponent is a function of $n$, i.e. of how carefully the 
average is done: on the bottom plot, we see how fast a good estimate of the 
average value is obtained as a function of $n$, for the two fixed widths 2 and
4. We see that at $n=20$, an accurate estimate of the average value has
been reached. For this ``true'' average $G_{20}(\Delta t)\simeq 
G_\infty(\Delta t)$, our estimate of the effective mass of the decay 
 $e^{-m_{eff}\Delta t/2}$  between $\Delta t =2$ and $\Delta t =4$  
is 1.22(2), and  1.77(20) between $\Delta t =4$ and $\Delta t =6$.
This is compatible with the mass of the lightest 
$2^{++}$ glueball we shall calculate below. 
Thus it  seems that, at large widths, $G_\infty(\Delta t)$ 
does decay with a mass close to the lightest state sharing the symmetry of 
our operator. The question now is, when has a good enough estimate of this 
quantity been achieved so that statistical errors are maximally reduced? 
 
%%%%%%%%%%%%%%%%%%%%%%%%%%%%%%%%%%%
\subsection{Optimisation procedure}
%%%%%%%%%%%%%%%%%%%%%%%%%%%%%%%%%%%
Since errors are reduced with the number of measurements as $1/\sqrt{n}$ at
 large $n$, we plot
$(G_n(\Delta t))^2 \times n$ and, at fixed $\Delta t$,  choose 
$n$ to minimise this quantity. Eventually  this quantity must
increase, because the ``error'' $G$ tends to a constant at large $n$.

We have already seen (fig. (\ref{stud}) right) that for  the smallest width,
 the average is reached very smoothly. As a consequence, the optimal $n_2$
according to our criterion is small: $n_2$=5. If we now fix this parameter
at that value and move to the width 4 block, 
 fig. (\ref{opti4}) reveals that the corresponding optimisation curve 
reaches its minimum between $n_4=70$ and 90. Indeed, $G$ reaches its 
asymptotic value around these values of $n_4$. The data was obtained as an 
average over 30 boundary conditions. Between submeasurements, 5 update
sweeps are performed.
%%%%%%%%%%%%%%%%%%%%%%%%%%%%%%%%%%%%%%%%%%%%%%%%%%%%%%%%%%%%%%%%%%%%%%%%%%%%%
\FIGURE[htb]{

\centerline{\begin{minipage}[c]{10cm}
    \psfig{file=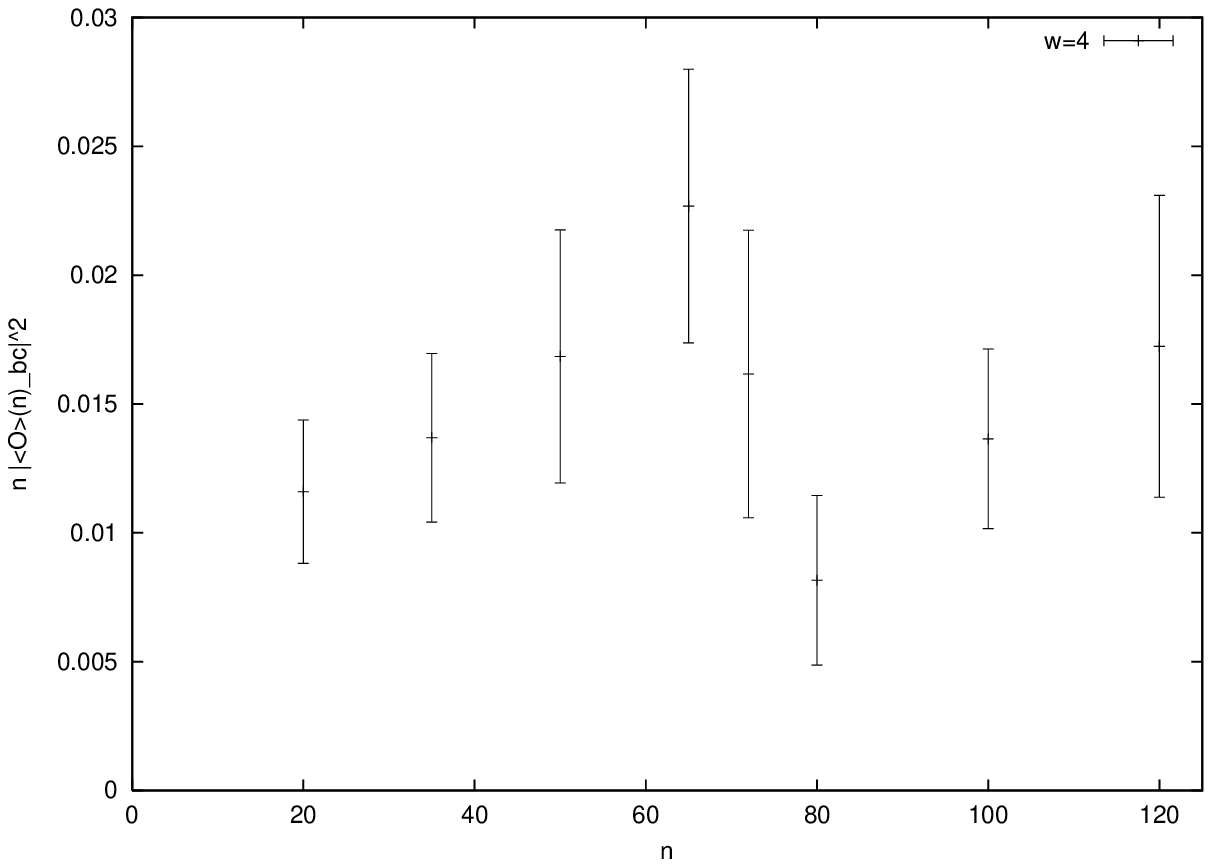,width=10cm}
    \end{minipage}}

\caption[a]{Optimisation curve for the width 4 level}

\la{opti4}
}
%%%%%%%%%%%%%%%%%%%%%%%%%%%%%%%%%%%%%%%%%%%%%%%%%%%%%%%%%%%%%%%%%%%%%%%%%%%%%
The error reduction achieved with slightly different parameters below
shows that the success of the algorithm
is not all too dependent on a fine-tuning of parameters.

Finally, we note that to optimise the parameters for a $0^{++}$ operator, whose
expectation value in the full lattice does not vanish, we need to minimise
the fluctuation around that value (which has to be known in advance).

%%%%%%%%%%%%%%%%%%%%%%%%%%%
\subsection{Results}
%%%%%%%%%%%%%%%%%%%%%%%%%%%
We compute the $0^{++}$ and $2^{++}$ correlation functions
formed with a $4\times 2$ rectangular Wilson loop  at 4 lattice spacings. 
We use two smearing steps on the operator. 
Errors are estimated with a standard jackknife analysis using 26 bins. 
A two-level scheme is implemented: 
the 8 time-slices are split into 2 time-blocks of width 4,  which are
in turn decomposed into 2 time-blocks of width 2. We restrict ourselves to
the measurement of the correlation function at even time-separations.

For the $0^{++}$ correlation function at 4 lattice spacings, 
one ``measurement'' comprises  
10 submeasurements at the lower level, 40 at the upper level. When
performing the latter, we are free to compute the 0 and 2 lattice spacing
 correlation in the standard way (thus the error reduction is only applied
to the $t=4a$ correlation).
%Between these subsmeasurements, 5 sweeps
%were done at the lower level and 10 at the upper level.
We collected 260 of these compound measurements. 

We proceed similarly in the $2^{++}$ case with following parameters:
 one ``measurement'' comprises  8 submeasurements
at the lower level, 150 at the upper level, 
each of these being preceded by 5 sweeps;
our program needs about 8.3 minutes on a standard alpha machine to perform
one of these compound measurements. We collected 520 of them; 
because they are time-consuming, we perform $\sim 200$ sweeps
between them to reduce their statistical dependence. 

The following values for the correlation functions, 
as well as their corresponding local effective masses 
(extracted from a cosh fit),  were obtained:
%\vspace{0.5cm}
\begin{center}
\begin{tabular}{|c|c|c||c|c|}
\hline
$t/a$ &  $\< {\cal O}_0^\dagger(0)~ {\cal O}_0(t)\>$&$am_{eff}^{(0)}(t)$  &$\< {\cal O}_2^\dagger(0)~ {\cal O}_2(t)\>$& $am_{eff}^{(2)}(t)$ \\
\hline
0   &  1.0000(65)&       &  $  1.0000(14)$ &  \\
1   &            &  1.017(35)   &     &  2.151(75)    \\
2   &   0.1331(99)&       &   $ 1.36(20) \times 10^{-2}$ &  \\
3   &            & 0.929(49)   &     &   1.794(74)   \\
4   &   0.0406(39)&      & $ 7.49(70)\times10^{-4}$&  \\
\hline
\end{tabular}
\end{center}
\vspace{0.5cm}
Fig. (\ref{corfun}) shows a plot of the two correlation functions. 
The small error bars appear to be roughly constant on the logarithmic scale.
The $t=4a$ correlation of the $2^{++}$ operator shows that a factor 20 error 
reduction has been achieved with respect to the $t=0$ point. It is already
at an accuracy that would be inaccessible on current single-processor
 machines with the standard algorithm. Indeed the latter yields  error bars 
that are roughly independent of $t$ and would have given an error comparable
to our error on the $t=0$ data, where we do not achieve any error reduction.  
 The naive error-reduction  estimate
of section~\ref{errestim} evaluates to $\exp(1.8\times4/2)\simeq36$. Not
unexpectedly, the observed reduction factor is somewhat 
lower than this naive estimate, presumably because the configurations
 generated at fixed boundary conditions are quite strongly correlated.

%%%%%%%%%%%%%%%%%%%%%%%%%%%%%%%%% FIGURE %%%%%%%%%%%%%%%%%%%%%%%%%%%%%%%%%%%
\FIGURE[htb]{

\centerline{~~\begin{minipage}[c]{13cm}
    \psfig{file=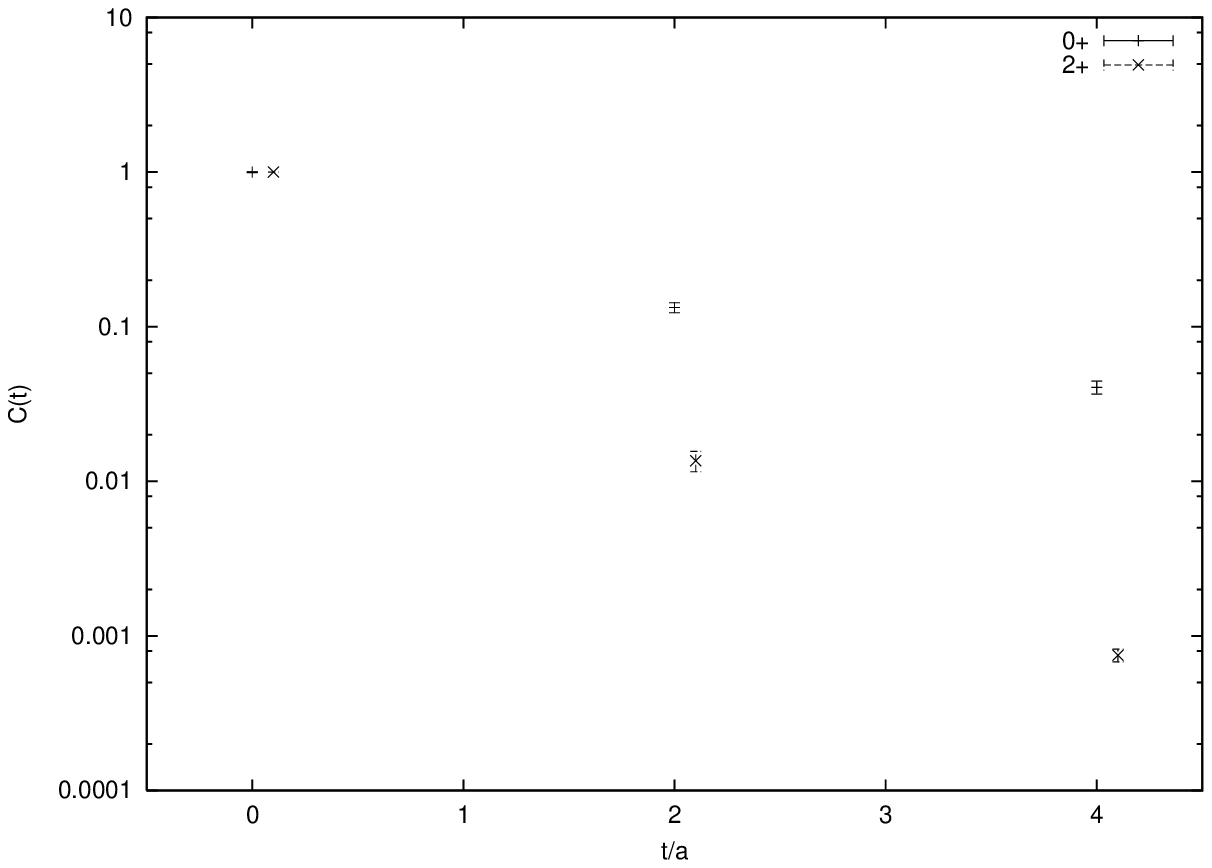,angle=0,width=13cm}
    \end{minipage}}%

\caption[a]{The $0^{++}$ and $2^{++}$ correlation functions}
\la{corfun}
}
%%%%%%%%%%%%%%%%%%%%%%%%%%%%%%%%%%%%%%%%%%%%%%%%%%%%%%%%%%%%%%%%%%%%%%%%%%%%

\section{Conclusion}
In this paper we have proposed a general scheme in which the accuracy of 
numerically computed $n$-point functions in local field theories could be 
improved by the use of nested averages. While the procedure is known to be 
exact, the efficiency of the algorithm must ultimately be tested on a 
case-by-case basis. However, a simple application to $SU(3)$ Wilson loop
correlations showed that, in some cases at least, the multilevel algorithm
drastically reduces statistical errors. 
It can straightforwardly be used in conjunction with the smearing and blocking
techniques. A further nice feature of the 2-point function
 case is that previous knowledge of the low-energy spectrum provides useful
guidance in the tuning of the algorithm's many parameters. Also, the error
reduction achieved was roughly as anticipated. We intend to use
multilevel algorithms to extract the higher-spin spectrum~\cite{hspin}
of the $SU(3)$ gauge theory, where the higher masses involved and the 
successful use of the variational method \cite{mart_var} require a high level
of accuracy on the correlation functions.

Independently of statistical errors, it is much harder to determine quantities
extracted from $n$-point functions with $n\geq 3$; if these difficulties
are overcome, the multilevel scheme could prove a decisive asset in those
 numerical measurements.

\acknowledgments{The author thanks M.~Teper for helpful discussions
and M.~L\"uscher for sharing his experience on nested Monte-Carlo algorithms.
He is also grateful to Lincoln College for its hospitality and the Berrow
Trust for  financial support.}

\end{document}